\documentclass[aps,prl,twocolumn,showpacs,superscriptaddress,groupedaddress]{revtex4-1}

\usepackage{graphicx}
\usepackage{dcolumn}
\usepackage{amssymb}
\usepackage{amsfonts}
\usepackage{amsmath}
\usepackage{bm}
\usepackage{color}
\usepackage{hyperref}
\usepackage[english]{babel}
\usepackage{version}

\begin{document}

\title{Vanishing Hall Response of Charged Fermions in a Transverse Magnetic Field}

\author{Michele Filippone}
\author{Charles-Edouard Bardyn}
\author{Sebastian Greschner}
\author{Thierry Giamarchi}
\affiliation{Department of Quantum Matter Physics, University of Geneva, 24 Quai Ernest-Ansermet, CH-1211 Geneva, Switzerland}


\begin{abstract}
We study the Hall response of quasi-two-dimensional lattice systems of charged fermions under a weak transverse magnetic field, in the ballistic coherent limit. We identify a setup in which this response vanishes over a wide range of parameters: the paradigmatic ``Landauer-B\"uttiker'' setup commonly studied for coherent quantum transport, consisting of a strip contacted to biased ideal reservoirs of charges. We show that the effect does not rely on particle-hole symmetry, and is robust to a variety of perturbations including variations of the transverse magnetic field, chemical potential, and temperature. We trace this robustness back to a topological property of the Fermi surface: the number of Fermi points (central charge) of the system. We argue that the mechanism responsible for the vanishing Hall response can operate both in noninteracting and interacting systems, which we verify in concrete examples using density-matrix renormalization group (DMRG) simulations.
\end{abstract}


\maketitle


Transport properties induced by electromagnetic fields are an active area of study in condensed matter physics. The Hall response, $\sigma_{\rm H}$, is of particular interest: It represents the off-diagonal response of a current density $\mathbf J$ to an electric field $\mathbf E$, $\sigma_{\rm H} = \varepsilon_{ij} \sigma_{ij}$, where $J_i = \sigma_{ij} E_j$, and $\varepsilon_{ij}$ is the Levi-Civita symbol. The Hall response
probes important geometric and topological properties of quantum systems: the Fermi-surface curvature of metals under weak magnetic fields~\cite{ong91_geometric_hall,tsuji58_hall_effect_cubic,haldane05_geometrical_hall_effect}, the Berry curvature of anomalous Hall systems~\cite{haldane2004__berry_anomalous_hall_effect}, and related topological invariants of band insulators (e.g., the band-integrated Berry curvature)~\cite{thouless1982_quantized_hall_conductance,niu1985_quantized_hall_topological}. Studies of $\sigma_{\rm H}$ are ubiquitous in fields focused on topological quantum matter~\cite{xiao2010_berry_phase_review} and synthetic realizations thereof~\cite{bloch2008_many_body_cold_atoms_review,*jotzu2014_haldane_model_cold_fermions,*mancini2015_chiral_edges_neutral_fermions,*tai2017_microscopy_HHmodel,*miyake2013_Harper_model_optical_lattices,*genkina2018_imaging,*jaksch2003_synthetic_gauge_field,haldane2008_photonic_crystals,*wang2009_topological_photonic_crystal,*hafezi2011_optical_delay_topological,*ningyuan2015_time_site_resolved_topo_circuit}.

Scattering is an essential ingredient in conventional studies of the Hall response: In the two-dimensional (2D) Hall effect~\cite{hall_1879_hall_effect}, e.g., Boltzmann-type approaches~\cite{ziman_book} allow us to reproduce the observed Hall constant, $R_{\rm H}$, in the limit of weak transverse magnetic fields $B$
: $R_{\rm H} \equiv -\sigma_{\rm H}/(\sigma_{xx} \sigma_{yy} B) \sim -1/(ne)$, where $n$ is the density of carriers with charge $e$, and $x$ $(y)$ denotes the longitudinal (transverse) direction. Scattering is also key to explaining the plateaus of quantized Hall conductance ($\sigma_{\rm H} = \nu e^2/h$ for filling factor $\nu$) appearing at stronger magnetic fields, in the quantum Hall regime~\cite{klitzing1980_quantum_hall,tsui1982_fractional_quantum_hall,bernevig2013_book}.

As ballistic quantum systems become more and more accessible in experiments~\cite{ella2018_simultaneous_voltage_current,bachmann2019_super_gemoetric_focusing,genkina2018_imaging}, new challenges are emerging for theory beyond Boltzmann-type approaches, despite significant efforts for mesoscopic systems~\cite{roukes1987_quench_hall_effect,*ford1988_quench_hall,baranger1991_classicla_quantum_ballistic,*beenakker1988_quench_hall,*kirczenow88_ballistic_hall_effect}. For example, $\sigma_{ii}$ can be infinite in clean interacting systems, even at finite temperature
~\cite{prosen2011_openxxz_ballistic,ljubotina2017_spin_diffusion_integrakble}. The connection between Hall response and carrier density is not even clear in the presence of interactions~\cite{hagen1990_anomalous_hall_superconductors,badoux2016_change_carrier_superconductor,smith1994_sign_reversal_hall}, remaining an important theoretical issue~\cite{kapitulnik2017_anomalous_metals_supra}. Recent progress was made with the calculation of $R_{\rm H}$ in dissipative metallic systems~\cite{auerbach2018_hall_number}, where $\sigma_{ii}$ is finite at zero frequency (in contrast to gapped systems where $\sigma_{ii} = 0$, and $\sigma_{\rm H}$ can be calculated in torus geometry~\cite{thouless1982_quantized_hall_conductance,niu1985_quantized_hall_topological,hatsugai1993_chern_number,avron1985_hall}). Nevertheless, the Hall response of coherent ballistic systems remains a largely unexplored field.

\begin{figure}[b]
    \centering
    \includegraphics[width = \linewidth]{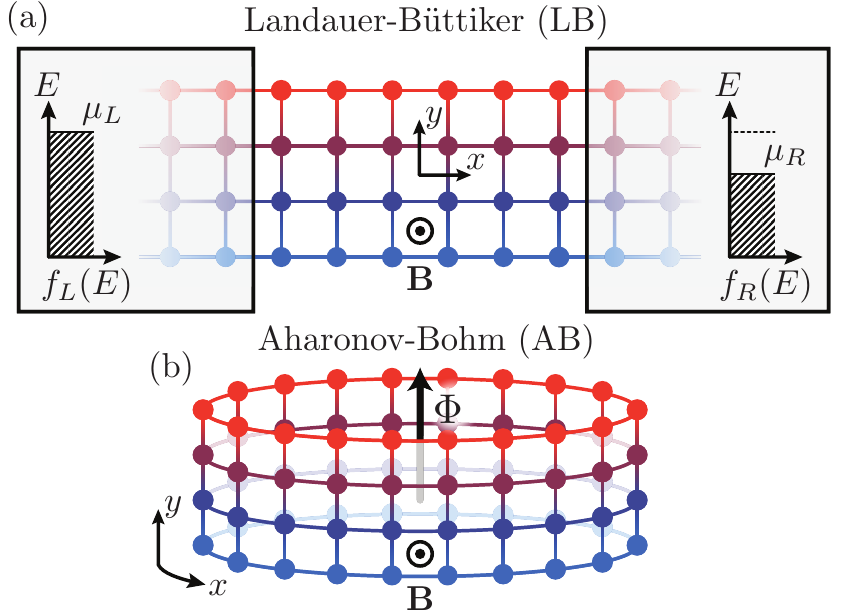}
    \caption{{\bf (a)} Schematic ``Landauer-B\"uttiker (LB)'' setup enabling a vanishing Hall response for charged fermions in a magnetic field $B$: A ballistic lattice system with a finite number of fermionic degrees of freedom (e.g., lattice sites) in the transverse ($y$) direction is connected to ideal reservoirs of charges (in gray) with weakly-biased chemical potentials $\mu_L$ and $\mu_R < \mu_L$ [corresponding to Fermi-Dirac distribution $f_L(E)$ and $f_R(E)$]. {\bf (b)} Ballistic Aharonov-Bohm (AB) setup where the Hall response is, in contrast, generically finite [and current in the longitudinal ($x$) direction is induced by a magnetic flux $\Phi$, instead of biased reservoirs].}
    \label{fig:setups}
\end{figure}

In this Letter, we identify a ballistic coherent setup in which charged fermions under a weak transverse magnetic field exhibit a strictly vanishing Hall response. We demonstrate this effect in noninteracting quasi-2D lattice systems at zero temperature, in a transport setup where the system is connected to weakly-biased ideal reservoirs of charges. We show that the Hall response remains suppressed under a wide variety of perturbations: variations of the magnetic field, chemical potential, temperature, and particle-hole symmetry breaking. We relate this remarkable robustness to the topological nature of a key property underpinning the effect: the number of Fermi points (central charge) of the system. We extend our results to interacting systems, demonstrating similar effects using DMRG.

{\it Hall response of ballistic systems ---}
%
We consider lattice systems in a quasi-2D geometry, i.e., with edges as well as a finite number of fermionic degrees of freedom (lattice sites, in particular) in the $y$ direction. Edges imply that the transverse current $J_y$ vanishes in the low-frequency limit $\omega \rightarrow 0$ of the longitudinal electric field $E_x$. In that case, the Hall response is described by the transverse polarization difference $\Delta P_y(x, t) = \int_{t_0}^t dt' J_y(x, t')$. We set the initial polarization $P_y(x, t_0)$ (at time $t_0$ right before applying $E_x$) to zero, which corresponds to a gauge choice~\cite{resta1992_theory_polarization,*resta1994_review_polarization,king_smith1993_theory_polarization,watanabe2018_inequivalent_berry_phases}, and denote $\Delta P_y(x, t) \equiv P_y(x, t)$.

The relation between $P_y$ and $\sigma_{\rm H}$ can be derived within standard linear response theory~\cite{kubo1957_kubo_formula,*greenwood1958_conduction_metals}: Expressing the electric field as $E_x = -\partial_t A_x$ (with $e = \hbar = c = 1$), without loss of generality, one can write
\begin{equation} \label{eq:Hrelation}
    P_y(k, \omega) = -\sigma_{\rm H}(k, \omega) A_x(k, \omega)\,,
\end{equation}
where $k$ is the crystal momentum along $x$. As we detail in the Supplemental Material (SM)~\cite{SM}, this can be seen as a Kubo formula for the transverse polarization induced by a time-dependent vector potential~\footnote{We refer to the function $A_x(x, t)$ in $E_x = -\partial_t A_x$ as a ``vector potential'', though it need not coincide with the magnetic vector potential.}, $P_y(x, t) = i\sum_{x'} \int dt' \theta(t-t') \langle [P_y(x, t), J_x(x', t')] \rangle A_x(x', t')$.

Equation~\eqref{eq:Hrelation} enables very {\it different} Hall responses $\sigma_{\rm H}$ depending on the nature of $E_x$, for the {\it same} longitudinal current $J_x$. Here we consider a paradigmatic setup for coherent quantum transport: a system with two ends in the $x$ direction, where $E_x$ (or $J_x$) is generated via ideal contacts to two external reservoir of charges (left and right) with chemical potentials $\mu_{\rm L}$ and $\mu_{\rm R}$ [Fig.~\ref{fig:setups}(a)].
%
%
In this ``Landauer-B\"uttiker'' (LB) setup~\footnote{Name chosen in connection with the transport formalism of the same name~\cite{landauer1970_scatt_theory,lesovik11_scattering_review}.}, $J_x$ is related to the potential difference $eV \equiv \mu_{\rm L} - \mu_{\rm R}$ via the conductance $G$ of the system, i.e., $J_x = GV$. Without interactions, the polarization $P_y^{\rm LB}$ can be calculated using conventional scattering theory~\cite{lesovik11_scattering_review}, with conductance $G$ derived from the Landauer formula. 
Kubo's formalism [Eq.~\eqref{eq:Hrelation}] provides an instructive equivalent approach~\cite{stone1988_kubo_landauer,*baranger1089_kubo_landauer_magnetic}. As we detail in the SM~\cite{SM}, the LB setup can be described by $A_x(x,t) = -V e^{-i\omega t} \delta(x)$, corresponding to a potential drop of amplitude $V$ at the position $x = 0$ of contact between the system and the left reservoir. Since $A_x(x,t)$ is local, the stationary transverse polarization $P_y^{\rm LB}$ takes the form of an {\it integral} of the Hall response {\it over all momenta}~\cite{SM}:
\begin{equation} \label{eq:PyLB}
    \frac{P_y^{\rm LB}}{J_x} = -G^{-1} \lim_{\omega\rightarrow 0} \frac{1}{2\pi} \int dk \, e^{ikx}\frac{\sigma_{\rm H}(k, \omega)}{\omega + i0^+}\,,
\end{equation}
where $i0^+$ is a small positive imaginary part.

\begin{figure}[t]
    \centering
    \includegraphics[width = \linewidth]{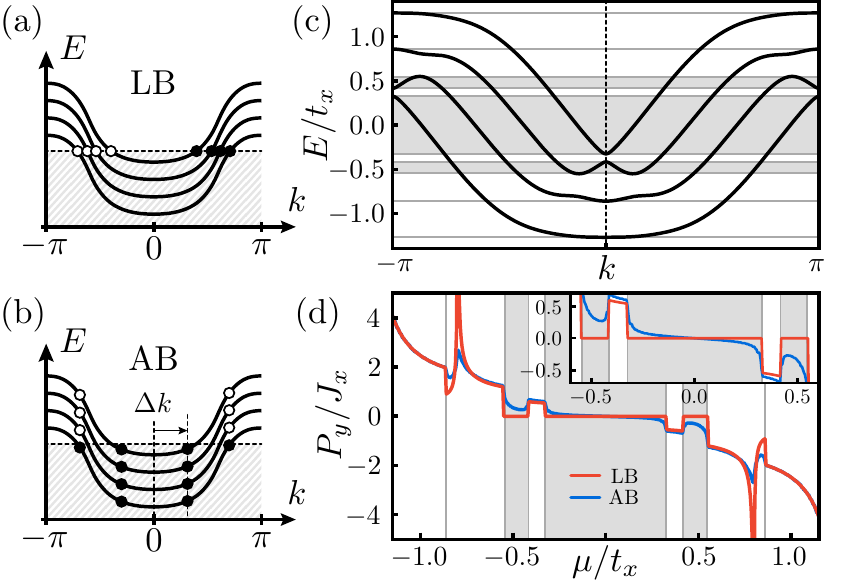}
    \caption{{\bf (a, b)} Schematic band structures showing the key single-particle states for the Hall response: In the LB setup, the current $J_x \neq 0$ is induced by occupied states (full dots) at $k > 0$ in a small energy window $[\mu_R, \mu_L]$ around the chemical potential $\mu$ (horizontal dashed line). Left-moving states in this window are empty. In the AB setup, in contrast, $J_x \neq 0$ is induced by the spectral flow $\Delta k = \Phi/N_x$ of \emph{all} states with threaded magnetic flux $\Phi$. {\bf (c)} Band structure of the HH model computed for $N_y = 4$, $B = 0.7$, and $t_y = 0.5 t_x$. Horizontal lines (dark gray) indicate energies at which the number $c(\mu)$ of Fermi points at $k > 0$ changes (by one), with $c(\mu) = N_y$ in shaded (light gray) regions. {\bf (d)} Hall response of the system in (c) in LB vs. AB setups. In regions where $c(\mu) = N_y$ [shaded regions as in (c)], $P_y/J_x$ strictly vanishes in the LB setup, while it only goes to zero at the particle-hole symmetric point $\mu = 0$, in the AB setup (see inset zoom).}
    \label{fig:bandStructureAndResponses}
\end{figure}


To illustrate how different the Hall responses of ballistic coherent systems can be, we investigate an additional ``Aharonov-Bohm'' (AB) setup: a contactless system forming a ring in the $x$ direction, where $J_x$ is induced by a time-dependent magnetic flux [Fig.~\ref{fig:setups}(b)]. In that case, $A_x(x,t)$ corresponds to the magnetic vector potential describing the flux, i.e., $A_x(x,t) = e^{i\omega t} \Phi/N_x$, where $N_x$ is the number of lattice sites in the $x$ direction. The flux induces a persistent current~\cite{buttiker1983_persistent_current,levy1990_persistent_currents,*kulik2010_review_persistent_currents,*saminadayar2004_equilibrium_mesoscopic,*bleszynski2009_persistent_currents}, $J_x = D \Phi/N_x$, where $D$ is the Drude weight~\cite{kohn64_drude,*shastry90_drude_weight,*millis1990_drude}, which in turn generates a {\it reactive} Hall response~\cite{prelovsek99_hall_correlated,*zotos00_reactive_hall,greschner2018_universal_hall_response} [see Fig.~\ref{fig:bandStructureAndResponses}(b)]. In contrast to Eq.~\eqref{eq:PyLB}, and in agreement with known transport results~\cite{luttinger1964_theory_transport_coefficients}, the stationary transverse polarization $P_y^{\rm AB}$ is then related to the {\it zero-momentum} component of $\sigma_{\rm H}$ alone~\cite{SM}:
\begin{equation} \label{eq:PyAB}
    \frac{P_y^{\rm AB}}{J_x} = -D^{-1} \lim_{\omega \rightarrow 0} \sigma_{\rm H}(0,\omega)\,.
\end{equation}

{\it Hall response in the LB setup ---}
%
We now detail the LB setup and derive an explicit formula for the response $P_y^{\rm LB}$ at zero temperature, in the low-bias limit $\mu_{\rm L} \rightarrow \mu_{\rm R} \equiv \mu$. We focus on weak magnetic fields $B \lesssim 1/N_y$, where $N_y$ is the number of lattice sites in the $y$ direction. This ensures that energy bands hybridize in a single Brillouin zone despite momentum shifts induced by the minimal coupling of system charges to $B$~\footnote{The system contains at most one magnetic unit cell in the $y$ direction.}. The resulting spectrum generically consists of $M_y$ bands, where $M_y$ is the number of fermionic degree of freedom (d.o.f) along $y$. We assume that $M_y = N_y$ (one d.o.f per lattice site), for simplicity, and consider systems whose energy spectrum is symmetric under momentum reversal $k \rightarrow -k$.

Our results apply to generic lattice models in this framework. For concreteness, however, we consider the Harper-Hofstadter (HH) model~\cite{harper55_harper_model,*hofstadter76_hofstadter_model} with Hamiltonian $H_{\rm HH} = -\sum_{x,y} [t_x e^{iBy} c^\dagger_{x,y} c_{x+1,y} + t_y c^\dagger_{x,y} c_{x,y+1}]/2 + \mbox{H.c}$ in the Landau gauge, where $t_x, t_y$ are hopping amplitudes and $c^\dagger_{x,y}$ creates a fermion on site $(x, y)$. The spectrum of $H_{\rm HH}$ is symmetric under $k \rightarrow -k$ by a combination of time reversal (TR) and spatial inversion in the $y$ direction. The corresponding effective TR symmetry is described by the operator $\Theta = I_y \mathcal K$, where $I_y$ permutes positions $y$ around the center of the system, and $\mathcal K$ describes complex conjugation. As $[H_{\rm HH}, \Theta] = 0$, the action of $\Theta$ on an eigenstate $|\psi_k(E)\rangle$ of $H_{\rm HH}$ with momentum $k$ and energy $E$ gives a (non-necessarily distinct~\footnote{Since $\Theta^2 = +1$ (due to the spinless nature of the fermions that we consider), Kramers' theorem does not hold.}) eigenstate $\Theta |\psi_{k'}(E)\rangle$ with $k' = -k$ and the same energy.

The Hall response can be derived using scattering theory~\footnote{Or, equivalently, by using Eq.~\eqref{eq:PyLB} (see SM~\cite{SM}).}: In the low-bias zero-temperature limit, the conductance is $G = G_0 \sum_j T_j$, where $G_0 = e^2/h = 1/(2\pi)$ is the conductance quantum, and $T_j$ is the transmission probability, at the chemical-potential energy $\mu$, of scattering modes $\psi_j(x, y)$ incoming from the left reservoir. We assume that the reservoirs are large (infinite) regions described by $H_{\rm HH}$ (with chemical potentials $\mu_L$ and $\mu_R$, respectively), so that scattering modes take a similar form as the system's eigenmodes. In that case, $T_j = 1$ for all modes $\psi_j(x, y)$ available at energy $\mu$. 
Relevant modes have the asymptotic form $\psi_j(x \rightarrow -\infty, y) = e^{ik_{F,j}x} w_j(y)/v_{F,j}$, where $k_{F,j}$ ($v_{F,j}$) denote the system's Fermi momenta (velocities), and $w_j(y)$ its transverse wavefunctions. The conductance is $G = c(\mu) G_0$, where $c(\mu)$ is the number of Fermi points at $k > 0$ [see Fig.~\ref{fig:bandStructureAndResponses}(a)]. As we detail in the SM~\cite{SM}, Eq.~\eqref{eq:PyLB} becomes
\begin{equation} \label{eq:pollandauer}
    \frac{P_y^{\rm LB}(\mu)}{J_x} = \frac{1}{c(\mu) G_0} \sum_{j = 1}^{c(\mu)} \sum_{y} y \frac{w_j(y)^2}{v_{F,j}} \,.
\end{equation}
More explicit expressions can be found in the SM for (i) finite $B$, $N_y = 2$, and (ii) $B \rightarrow 0$, $N_y \geq 2$.

We used the quantum-transport simulation package ``Kwant''~\cite{groth14_kwant} to verify the above analytical results, to compute $P_y^{\rm LB}$ for arbitrary $B$ and $N_y$, and to compare $P_y^{\rm LB}$ to $P_y^{\rm AB}$. Our results, illustrated in Fig.~\ref{fig:bandStructureAndResponses}(b), demonstrate two key points: First and foremost, $P_y^{\rm LB}$ {\it vanishes identically} whenever $c(\mu) = N_y$, i.e., whenever the chemical potential $\mu$ crosses $N_y$ times the system's energy bands at $k > 0$. This is in stark contrast to what may be expected of particles with charges of the same sign in a finite magnetic field. Provided that $c(\mu) = N_y$, the Hall response remains zero irrespective of particle-hole symmetry (generically absent here), and of the specific values of $\mu$ and $B$ [see the region around $\mu/t_x = \pm 0.5$, e.g., in Fig.~\ref{fig:bandStructureAndResponses}(b)]. Second, the responses $P_y^{\rm LB}$ to $P_y^{\rm AB}$ are strikingly different, which reflects the clear differences between Eqs.~\eqref{eq:PyLB} and~\eqref{eq:PyAB}. Intuitively, this comes from the fact that the stationary states found in the LB and AB setups are different [Fig.~\ref{fig:bandStructureAndResponses}(a,~b)], with distinct polarizations $P_y$, though they carry the same current $J_x$. The two responses only coincide when $c(\mu) = 1$, which can be understood from simple analytical considerations~\cite{SM}.

{\it Topological origin of the vanishing Hall response ---}
%
We now demonstrate that $P_y^{\rm LB} = 0$ arises from two elements: (i) the topological nature of $c(\mu)$, and (ii) the traceless nature of the operator $\hat P_y$ describing the polarization. The number $c(\mu)$ of Fermi points with $k > 0$ is topological in the sense that it coincides with the {\it central charge} of the system---the number of gapless modes at $k > 0$, if we interpret the system as a Luttinger liquid (without interactions, for now). The polarization operator is $\hat P_y = e Y$, where $Y = \sum_{x, y} y c_{x, y}^\dagger c_{x, y}$ describes the ``center-of-mass'' position along $y$. To ensure that $\langle \psi_i^{\rm LB}(\mu) | \hat P_y | \psi_i^{\rm LB}(\mu) \rangle = 0$ in the initial state $| \psi_i^{\rm LB}(\mu) \rangle \equiv | \psi_i \rangle$ with zero bias ($V, J_x = 0$), corresponding to our gauge choice for the polarization, we set $y = 0$ at the center of the system. The operator $\hat P_y$ then satisfies $I_y^T \hat P_y I_y = -\hat P_y$. It is traceless, and $P_y^{\rm LB} = 0$ at $V = 0$ is enforced by the symmetry between $k$ and $-k$: Indeed, $| \psi_i \rangle$ is the many-body ground state of $H_{\rm HH}$ with single-particle states occupied symmetrically around $k = 0$ up to the chemical potential $\mu$. It is symmetric under $\Theta$ (i.e., $\Theta | \psi_i \rangle = \pm | \psi_i \rangle$), such that $\langle \psi_i | \hat P_y | \psi_i \rangle = \langle \psi_i | \Theta^\dagger \hat P_y \Theta | \psi_i \rangle = \langle \psi_i | I_y^T \hat P_y I_y | \psi_i \rangle = - \langle \psi_i | \hat P_y | \psi_i \rangle$.

Upon applying a finite bias $V \neq 0$ to generate a stationary current $J_x$ in the ``final'' state $| \psi_f^{\rm LB}(\mu) \rangle \equiv | \psi_f \rangle$, the symmetry $\Theta$ breaks: The state $| \psi_f \rangle$ is a many-body stationary state consisting of single-particle states occupied symmetrically around $k = 0$ {\it except} at the chemical potential $\mu$ where single-particle states are occupied at $k > 0$ only. By symmetry, noncancelling contributions to the polarization must come from these $c(\mu)$ Fermi points. We index the latter with $j = 1, 2, \ldots, c(\mu)$, and denote by $|j\rangle$ the corresponding single-particle states ($|j\rangle \equiv |k_{F,j},s_j\rangle$ here, where $k_{F,j}$ and $s_j$ are the Fermi momentum and band index of the Fermi point $j$). The polarization takes the generic form
\begin{equation} \label{eq:PyLBFermiPoints}
    P_y^{\rm LB}(\mu) = \sum_{j=1}^{c(\mu)} n_j \langle j | \hat P_y | j \rangle,
\end{equation}
where $n_j = 1$ is the stationary occupation of $|j\rangle$.

We are now in position to show that $P_y^{\rm LB}(\mu)$ vanishes in a robust way whenever $c(\mu) = N_y$: The states $|j\rangle$ in Eq.~\eqref{eq:PyLBFermiPoints} belong to the eigenspace of $H_{\rm HH}$ with energy $\mu$, and are characterized by distinct momenta. Since they are not related by any symmetry~\footnote{The symmetry $\Theta$ relates states with {\it opposite} momenta.}, they form a basis for a Hilbert (sub)space of dimension $c(\mu)$. Consequently, when $c(\mu) = N_y$, Eq.~\eqref{eq:PyLBFermiPoints} reduces to
\begin{equation} \label{eq:PyLBFermiPoints2}
    P_y^{\rm LB}(\mu)|_{c(\mu) = N_y} = \sum_{j=1}^{N_y} \langle j | \hat P_y | j \rangle = \mathrm{Tr} \hat P_y = 0 \,.
\end{equation}
This key result represents a conservation law for the Hall response of the LB setup, which differs from the response of the AB setup, e.g., where {\it all} occupied single-particles states contribute to $J_x$ and $P_y$ (see Fig.~\ref{fig:bandStructureAndResponses}(b) and SM~\cite{SM}). Note that other potentially observable conservation laws can be derived from the tracelessness of $\hat P_y$ in bases of dimension $N_y$: In particular, replacing the set $\{ |j\rangle \}$ by a basis of Bloch eigenstates $\{ | k, s \rangle \}$ (with momentum $k$ and band index $s = 1, 2, \ldots, N_y$), one finds
\begin{equation} \label{eq:conservationLawFixedk}
    P_y(k) \equiv \sum_{s=1}^{N_y} \langle k, s | \hat P_y | k, s \rangle = 0 \,,
\end{equation}
implying that the transverse polarization (Hall response) of a system with $N_y$ bands vanishes in any momentum sector $k$ where all bands are equally occupied. This conservation law is directly related to the known zero-sum rule for the Berry curvature of all eigenstates of a Hamiltonian~\footnote{The quantity \unexpanded{$\langle k, s | \hat P_y | k, s \rangle$} can be seen as a Berry connection in the continuum limit where the position operator $Y = \hat P_y$ is represented by $Y = -i\partial_{k_y}$; see also Ref.~\cite{xiao2010_berry_phase_review}.}. The conservation law in Eq.~\eqref{eq:PyLBFermiPoints2} can be seen as an analog with fixed energy (and $k > 0$), instead of $k$.

{\it Robustness to perturbations ---}
%
The vanishing of $P_y^{\rm LB}$ for $c(\mu) = N_y$ is protected against temperature by an energy gap: the energy $\Delta \mu$ corresponding to the smallest chemical-potential variation required for $c(\mu) \neq N_y$. The Hall response is suppressed as $e^{-\beta |\Delta \mu|}$ at finite temperature $T = 1/\beta > 0$ (setting $k_B = 1$), accordingly, as we illustrate in the SM in an example with $N_y = 2$~\cite{SM}.

We remark that deviations from a strictly vanishing Hall response are also expected in the presence of disorder, as the effect relies on the ballistic coherent nature of the system~\footnote{Generic disorder also breaks the symmetry $\Theta$ connecting momentum sectors $k$ and -$k$.}. Disorder in quasi-1D systems generally leads to Anderson localization~\cite{abrahams79_anderson_localization,*abrahams10_anderson_localization,*lagendijk09_anderson_localization}. Nevertheless, provided that the scattering region connecting the two reservoirs is shorter than the corresponding localization length (scaling as $N_y t_x^2/W^2$ with disorder strength $W$~\cite{kappus1981_anomaly_anderson}; see SM~\cite{SM}), disorder can be regarded as a weak perturbation. In this regime, deviations of the disorder-averaged polarization $\langle P_y \rangle$ from zero scale as $W^2/t_x^2$~\cite{SM}, with large fluctuations around the average (as for typical conductance fluctuations in disordered systems~\cite{altshuler1985_conductance_fluctuations,*lee1985_universal_conductance_fluctuations}).

{\it Generalization to interacting systems ---}
%
Equation~\eqref{eq:PyLBFermiPoints2} applies whenever $c = N_y$ independent and equally occupied fermionic d.o.f (the Fermi points at $k > 0$ discussed so far) are responsible for the current $J_x \neq 0$. The noninteracting nature of the underlying many-body state is irrelevant. To demonstrate that our results extend to interacting systems with $c = N_y$, we consider the above HH model with $N_y = 2$ (two-leg ladder) and additional intra- and inter-leg interactions described by Hamiltonian terms $U_\parallel \sum_{x, y = \pm 1} n_{x, y} n_{x+1, y} + U_\perp \sum_x n_{x, 1} n_{x, -1}$, where $n_{x, y}$ is the density on site $(x, y)$.

\begin{figure}[t]
    \centering
    \includegraphics[width=1\linewidth]{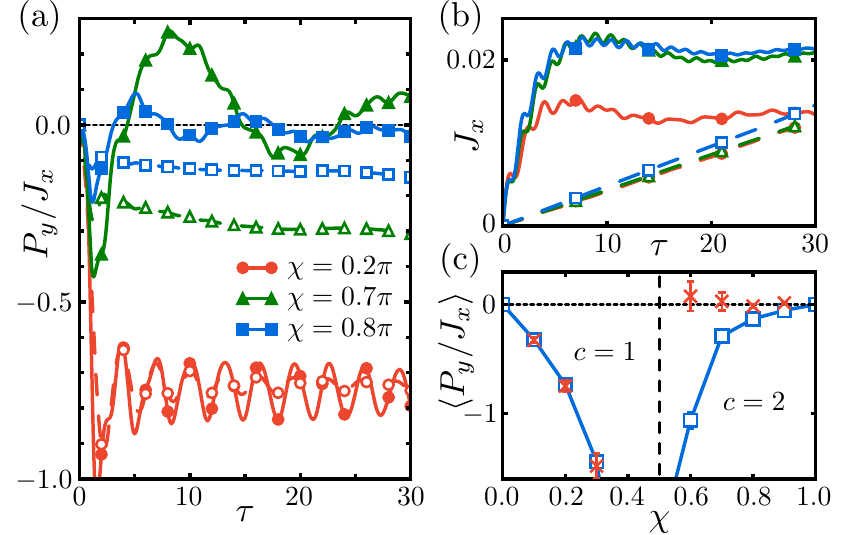}
    \caption{Numerical estimates of the Hall response of interacting fermions in the LB and AB setups (tDMRG simulations of the interacting HH model presented in the text, with $t_x = t_y = 1$, $U_\parallel = U_\perp = 1/2$, and $\epsilon = 0.01$, for $10$ particles in a full system of length $L_x = 60$).
    {\bf (a)} Time evolution of $P_y^{\rm LB}/J_x$ (filled symbols) and $P_y^{\rm AB}/J_x$ (empty symbols) for a total magnetic flux $\chi \equiv B/N_x = 0.2\pi$ (Luttinger-liquid phase with central charge $c = 1$), and $0.7\pi$ and $0.8\pi$ ($c = 2$). Lines interpolate a finer set of data points than shown symbols.
    {\bf (b)} Time evolution of $J_x$ alone for the same parameters as in (a).
    {\bf (c)} Average of $P_y^{\rm LB}/J_x$ ($\times$) and $P_y^{\rm AB}/J_x$ ($\Box$) over times $10 < \tau < 30$. The dashed line indicates the estimated transition between $c = 1$ and $c = 2$ phases. Averages coincide for $\chi = 0.4\pi$, while no stationary regime was reached for $0.5\pi$.}
    \label{fig:numerics}
\end{figure}

To simulate stationary transport conditions in the LB setup, we compute the time evolution of the full system with reservoirs described by a quenched steplike potential $-\epsilon \sum_{x < L_{\rm res}, y} n_{x, y} + \epsilon \sum_{x > L_{\rm sys} + L_{\rm res}, y} n_{x, y}$, where $L_{\rm sys/res}$ denotes the length of the system/reservoirs. We first set $\epsilon = 0$ and prepare the full system in its ground state using DMRG~\cite{White1992,Schollwoeck2011}. We then set $\epsilon > 0$, at time $\tau = 0$, and compute its evolution using tDMRG~\cite{Schollwoeck2011} and the ITensor library~\cite{ITensor}. We choose $L_{\rm sys} = 2$, for simplicity, and compute the Hall response $P_y^{\rm LB}/J_x$ in the middle of the system at times $1 \lesssim \tau/t_x \lesssim L_{res}$~\cite{Einhellinger2012}, averaging over a time window where $J_x$ is approximately stationary. Figure~\ref{fig:numerics} illustrates typical results for $U_\parallel = U_\perp = t_x/2$ and different magnetic fields. For comparison, we simulate transport in the AB setup by quenching, instead, a small linear potential $-(\epsilon/N_x) \sum_{x, y} x n_{x, y}$. While $J_x$ increases linearly in time in that case [Fig.~\ref{fig:numerics}(b)], the ratio $P_y^{\rm AB}/J_x$ oscillates around a constant value corresponding to the stationary Hall response~\cite{greschner2018_universal_hall_response}.

The results presented in Fig.~\ref{fig:numerics} are consistent with our theoretical expectations: First, the Hall responses found in the LB and AB setups are identical (for time averages, within errorbars) when the ground state prepared at $\tau = 0$ is a phase with central charge $c = 1$~\cite{SM,Holzhey1994,*Korepin2004,*Calabrese2004}. Second, the two responses differ completely when the system enters a $c = 2 = N_y$ phase, at some larger magnetic field [Fig.~\ref{fig:numerics}(c)]: While $P_y^{\rm LB}/J_x$ shows large oscillations around an average value consistent with $P_y^{\rm LB}/J_x = 0$, the response $P_y^{\rm AB}/J_x$ is finite. Along with additional data presented in the SM~\cite{SM}, our results are fully consistent with our theoretical arguments that the Hall response of the LB setup vanishes when $c = 2 = N_y$.

{\it Discussion ---}
%
Our results exemplify the rich and sometimes counterintuitive quantum transport phenomena that can occur in the ballistic coherent regime. The effect discovered here could be observed in solid-state or synthetic-matter experiments~\cite{ella2018_simultaneous_voltage_current,bachmann2019_super_gemoetric_focusing,genkina2018_imaging}. In fact, an experimental platform for the realization of the LB setup has recently been proposed~\cite{salerno2018_quantized_hall}. We emphasize that our results equally apply to bosons: In photonic systems~\cite{carusotto_2013,*hafezi2011robust,*kruk2017,*bellec2013,*poddubny2014,*downing2017}, for example, the conservation law found in this work could be observed by selectively populating the $c=N_y$ states responsible for the effect~\cite{Bardyn_2014}.

Our results could lead to additional clues towards a better understanding of the Hall response of strongly-correlated (non-Fermi-liquid) systems, for which low-energy-quasiparticle descriptions of quantum transport inexorably fail. More presently, they raise important questions regarding the behavior of the transverse polarization $P_y$ of interacting systems at finite temperatures: Although a transition to dissipative/metallic regimes is expected in such systems, explicit calculations of $P_y$ remain challenging~\cite{auerbach2018_hall_number}. Recent studies have shown the persistence of ballistic and superdiffusive regimes in specific cases~\cite{prosen2011_openxxz_ballistic,ljubotina2017_spin_diffusion_integrakble}. It will be interesting to investigate whether analogs exist in quasi-1D lattice systems.

{\it Acknowledgments ---}
%
We thank Jean-Philippe Brantut, Nigel Cooper, and Nathan Goldman for fruitful discussions, and acknowledge support by the Swiss National Science Foundation (FNS/SNF) under Division II. M.F. also acknowledges support from the FNS/SNF Ambizione Grant PZ00P2\_174038.

\bibliographystyle{apsrev4-1}
\bibliography{bibliography}

\end{document}